\documentclass[12pt]{iopart}
\usepackage{xcolor}
\usepackage{graphicx}
\expandafter\let\csname equation*\endcsname\relax
\expandafter\let\csname endequation*\endcsname\relax
\usepackage{amsmath,amssymb,amsfonts}
\DeclareMathOperator{\rea}{Re}

\begin{document}

\title{Zeta function regularization technique in the electrostatics context for discrete charge distributions}

\author{F. Escalante}

\address{Departamento de Física, Universidad Católica del Norte, Avenida Angamos 0610, Casilla 1280, Antofagasta, Chile.}
\ead{fescalante@ucn.cl}
\vspace{10pt}

\begin{abstract}
Spectral functions, such as the zeta functions, are widely used in Quantum Field Theory to calculate physical quantities. In this work, we compute the electrostatic potential and field due to an infinite discrete distribution of point charges, using the zeta function regularization technique. This method allows us to remove the infinities that appear in the resulting expression. We found that the asymptotic behavior dependence of the potential and field is similar to the cases of continuous charge distribution. Finally, this exercise can be useful for graduate students to explore spectral and special functions.
\end{abstract}

\maketitle
\section{Introduction}
The mathematician Leonard Euler introduced a version of the zeta function in 1737. However, it was extensively studied by the German mathematician Bernard Riemann, who, in his article ``On the Number of Primes Less Than a Given Magnitude'', published in 1859, took as a starting point the Euler's theorem (1737), where he proved that the sum of the reciprocals of the prime numbers is a divergent series \cite{edwards}. Riemann defined a function in the complex plane $\zeta(s)=\sum_{n=1}^{\infty}n^{-s}$, where $s\in\mathbb{C}$, and he proved that this sum converges when $\rea(s)>1$ and diverges when $\rea(s)=1$, which means that the $\zeta$-function has a simple pole.

The zeta functions properties, like Hurwitz, Barnes, and Epstein zeta functions, have been well-studied \cite{kirsten1}, and it is of our particular interest the Epstein zeta function \cite{epstein1, epstein2}, associated with sums of squares of integers. Besides, the zeta function has proved to be a very powerful tool in Quantum Field Theory (QFT), e.g. the Casimir Effect or theory of strong interactions \cite{barone, escalante, blau}.

One of the fundamental aspects of QFT is the task to extract results physically significant from quantities that, in principle, are ill-defined. Usually, these quantities are related to sequences of numbers $\lambda_{k}$ where $k\in\mathbb{N}$, which, for most applications, are eigenvalues of certain operators Laplace like that grow without bound when $k\to\infty$. In principle, the divergences that appear may be related to measurable physical quantities. Therefore, one must make sense of the calculations of those quantities. This can be achieved by a method called ``regularization'', where it extracts the relevant information the calculations made. This procedure generally leads to the appearance of a scale parameter, say $\Lambda$, which we can roughly say, allows giving a finite value to originally divergent expressions. This parameter is usually called ``cutoff'' or ``regulator''.
 
The original procedure in QFT consists of taking the limit $\Lambda\to\infty$ when the problematic (divergent) terms are isolated and suppressed from the desired expression. So the regulator has no relevant meaning and is just part of a mathematical artifact to make sense of certain formal expressions. It is important to note that there are various regularization methods, such as Pauli-Villars, dimensional regularization, cut-off regularization, among others \cite{george}.
 
However, these regularization schemes are not QFT exclusively. In the context of electrostatics, we can find situations where we have to deal with diverging quantities. Some authors have been studied the calculation of the electrostatic potential and field in continuous charge distributions, such as the infinite charged wire and the infinite charged plate, using cut-off and dimensional regularizations and renormalization schemes\cite{olness,hans,mundarain}.
 
This work aims to calculate the electrostatic potential and electric field of infinite one and two-dimensional lattices as an example of the application of the zeta function regularization technique, in the context of electrostatics.

\section{Point charges distribution}
The electrostatic potential due to a set of point charges is given by \cite{sears}

\begin{equation}
    \label{pot1}
    V=\frac{1}{4\pi\epsilon_0}\sum_{i}\frac{q_i}{r_i}
\end{equation}

\noindent where $r_i$ is the distance form the $i$th charge, $q_i$, to the point at which $V$ is evaluated.

\begin{figure}[h]
    \centering
    \includegraphics[width=0.4\linewidth]{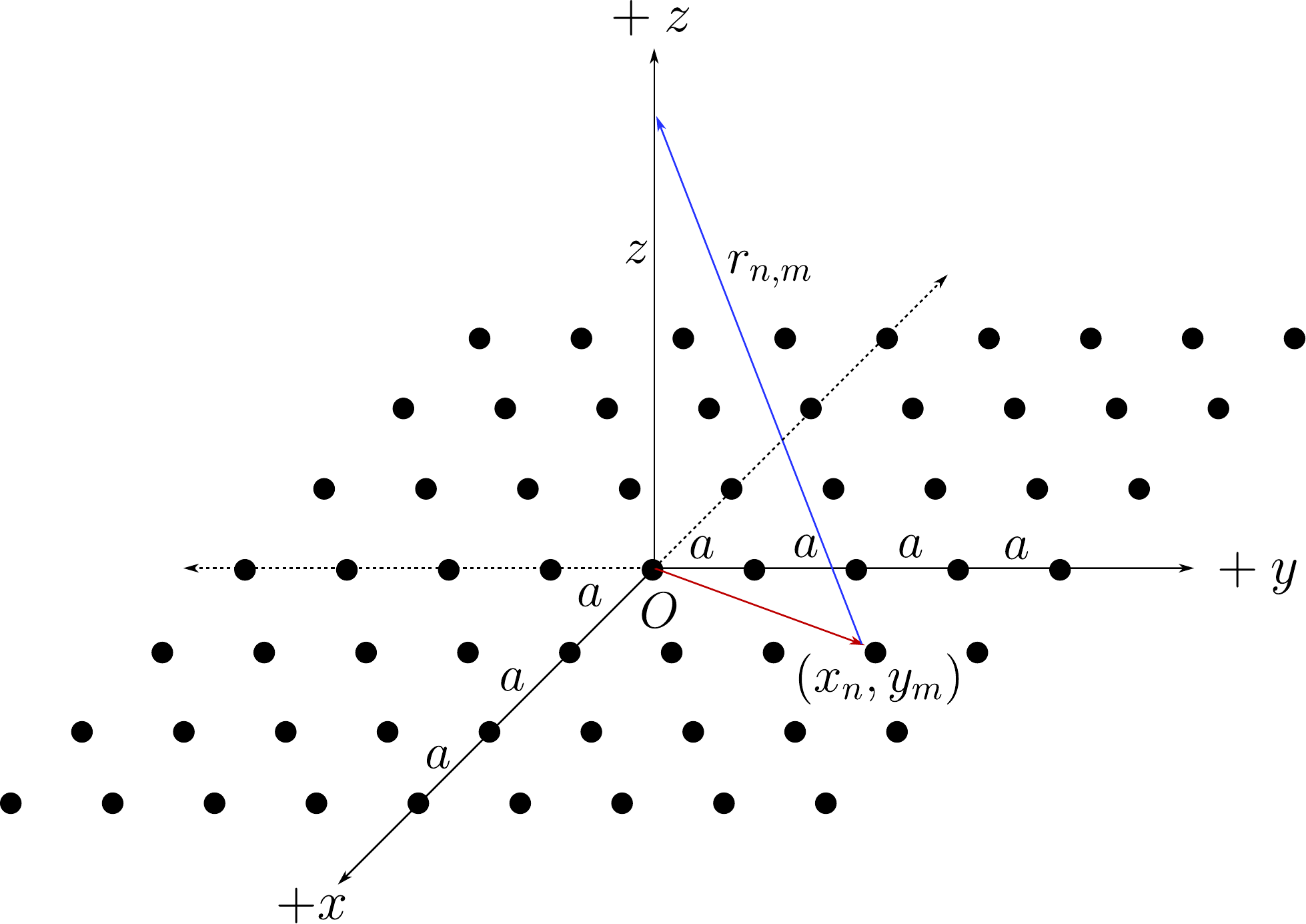}
    \caption{Graphic representation of the two dimensional point charge distribution.}
    \label{lattice}
\end{figure}

We consider a lattice of infinite identical point charges, located on the $xy$ plane, separated by a distance $a$ from each other in the $x$
and $y$ coordinates, as shown in figure (\ref{lattice}).

The origin of the coordinate system, $O$, is at the center of the lattice, where a point charge is located. The electrostatic potential on the $+z$ axis can be written as

\begin{equation}
    \label{pot2}
    V(z)=\frac{q}{4\pi\epsilon_0}\sum_{n,m=-\infty}^{\infty}\frac{1}{\sqrt{x_n^2+y_m^2+z^2}},
\end{equation}

\noindent where $x_n = an$ and $y_m = am$, $n$ and $m$ being integers, represent, respectively,
the $x$ and $y$ coordinates of point charges from the lattice. So we can rewrite the electrostatic potential as

\begin{equation}
    \label{pot3}
    V(\xi)=\frac{1}{4\pi\epsilon_0}\frac{q}{a}\sum_{n,m=-\infty}^{\infty}\frac{1}{\sqrt{n^2+m^2+\xi^2}},
\end{equation}

\noindent with $\xi = z/a$ as a dimensionless parameter, related to the $z$ coordinate. To obtain a dimensionless expression, we normalize
equation (\ref{pot3}) to the potential of a point charge at a distance a from the
origin along the $+z$ axis, $V_0=q/4\pi\epsilon_{0}a$. Thus, we get the expression

\begin{equation}
    \label{pott}
    \Tilde{V}(\xi)=\sum_{n,m=-\infty}^{\infty}\frac{1}{\sqrt{n^2+m^2+\xi^2}}.
\end{equation}

The sum in equation (\ref{pott}) diverges due to the infinite
summation range, and thus this divergence is analogous to the ultraviolet
divergence in QFT. On the other hand, if $\xi\to 0$, the sum also diverges at
the origine (when $n=m=0$), and such kind of divergence is analogous to the
infrared divergence in QFT \cite{mcomb}. However, in this work we are interested in the large $\xi$ regime. Therefore, in order to calculate the infinite sum, we must use an appropriate regularization method. For this purpose, we will use the zeta function method.

\section{Zeta function regularization method}
\label{metodo}
The multidimensional Epstein zeta function \cite{elizalde2} is defined by

\begin{equation}
  \label{zeta1}
  \zeta_{E}(s,d;\xi^2)=\sum_{n_{d}\in \mathbb{Z}}(n_1^2+n_2^2+\cdots+n_d^2+\xi^2)^{-s}
\end{equation}

\noindent where $d$ is the dimension of the function and $s\in \mathbb{C}$. By means of the Mellin transform, the multidimensional Epstein zeta function is given by
\begin{equation}
    \label{mellin}
    \begin{split}
    \zeta_{E}(s,d;\xi^2) & =\frac{1}{\Gamma(s)}\int_{0}^{\infty}dt\;t^{s-1}\sum_{n_{d}\in \mathbb{Z}}e^{-t(n_1^2+n_2^2+\cdots+n_d^2+\xi^2)}\\
    & =\frac{1}{\Gamma(s)}\int_{0}^{\infty}dt\;t^{s-1}e^{-t\xi^2}\sum_{n_{d}\in \mathbb{Z}}e^{-t(n_1^2+n_2^2+\cdots+n_d^2)}.
    \end{split}
\end{equation}

\noindent valid for $\rea(s)>d/2$. However, we are interested in the values where $\rea(s)<d/2$, so we need to make an analytical
continuation in the complex plane of $s$ to include the value $s=1/2$. Using the transformation
formula of the Jacobi theta function of a lattice \cite{conway}

\begin{equation}
    \label{jacobi}
\Theta(\omega)=\left(\frac{i}{\omega}\right)^{d/2}\Theta\left(-\frac{1}{\omega}\right),
\end{equation}

\noindent where $\Theta(\omega)=\sum_{n\in\mathbb{Z}^{d}}e^{-i\pi \omega(n_j\cdot n_j)}$, and with $\omega=it/\pi$, we have

\begin{equation}
    \label{jacobi2}
    \sum_{n\in\mathbb{Z}^{d}}e^{-t(n_j\cdot n_j)}=\left(\frac{\pi}{t}\right)^{d/2}\sum_{n\in\mathbb{Z}^{d}}e^{-\frac{\pi^{2}}{t}(n_j\cdot n_j)}.
\end{equation}

With the equation (\ref{jacobi2}), the equation (\ref{mellin}) can be written as it follows

\begin{equation}
    \label{mellin2}
    \zeta_{E}(s,d;\xi^2)=\frac{1}{\Gamma(s)}\int_{0}^{\infty} dt\;t^{s-1}e^{-t\xi^2}\left(\frac{\pi}{t}\right)^{d}\sum_{n\in\mathbb{Z}^{d}}e^{-\frac{\pi^2}{t}(n_1^2+n_2^2+\cdots+n_d^2)}.
\end{equation}

The term $n=0$ can be treated separately from the sum, so we obtain

\begin{equation}
    \label{mellin3}
    \begin{split}
    \zeta_{E}(s,d;\xi^2)=\frac{1}{\Gamma(s)}&\left\{\int_{0}^{\infty}dt\;t^{s-1}e^{-t\xi^2}\left(\frac{\pi}{t}\right)^{d}+\sum_{n\in\mathbb{Z}^{d}-\{0\}}\int_{0}^{\infty}dt\;t^{s-1}e^{-t\xi^2}\left(\frac{\pi}{t}\right)^{d}\right.\nonumber\\
    &\left.\times e^{-\frac{\pi^2}{t}(n_1^2+n_2^2+\cdots+n_d^2)}\right\}.
    \end{split}
\end{equation}

From the first integral we can recognize the Gamma function

\begin{equation}
    \label{gamma}
    \int_{0}^{\infty}d\tau\;\tau^{\alpha-1}e^{-A\tau}=\Gamma(\alpha)A^{-\alpha},
\end{equation}

\noindent in the second term, we use the integral representation for the modified Bessel function of second kind

\begin{equation}
    \label{bessel}
    \int_{0}^{\infty}d\tau\tau^{\alpha-1}e^{-\beta/\tau-\gamma\tau}=2\left(\frac{\beta}{\gamma}\right)^{\alpha/2}K_{\alpha}(2\sqrt{\beta\gamma}).
\end{equation}

Finally, we obtain an regularized expression for the multidimensional Epstein zeta function valid in whole the complex plane \cite{kirsten1}

\begin{equation}
    \label{zetafinal}
    \begin{split}
    \zeta_{E}(s,d;\xi^2)&=\frac{\pi^{d/2}}{\Gamma(s)}\frac{\Gamma(s-d/2)}{\xi^{2(s-d/2)}}+2\frac{\pi^s}{\Gamma(s)}\frac{1}{\xi^{s-d/2}}\sum_{n\in\mathbb{Z}^{d}-\{0\}}(n_1^2+n_2^2+\cdots+n_d^2)^{(s-d/2)/2}\\
    & \times K_{s-d/2}\left(2\pi\xi\sqrt{n_1^2+n_2^2+\cdots+n_d^2}\right).
    \end{split}
\end{equation}

The expression (\ref{zetafinal}) is analytic in the whole complex plane except for certain values for $d$. In general, when $d$ is even, the poles are located at $s=\{\frac{d}{2},\frac{d}{2}-1,...,1\}$, whereas when $d$ is odd, the poles are located at $s=\{\frac{d}{2},\frac{d}{2}-1,...,\frac{1}{2}, -\frac{2l+1}{2}\}$ where $l\in \mathbb{N}_{0}$. In other words, the poles are determined by the poles of the Gamma function of the first term of the expression (\ref{zetafinal}), where the zeta function is infinite and coincides with the original sum (\ref{pot3}).

\section{Electrostatic potential and the field of a one-dimensional lattice}

\begin{figure}[h]
    \centering
    \includegraphics[width=0.4\linewidth]{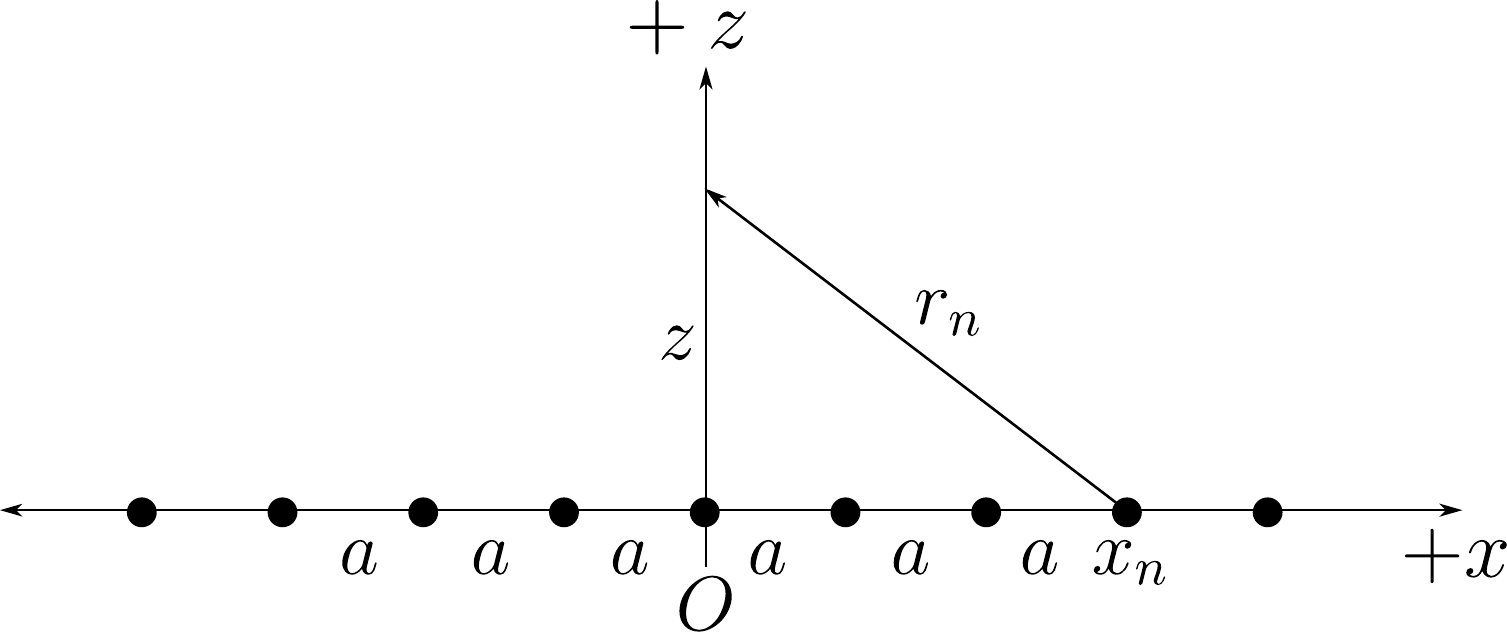}
    \caption{Graphic representation of the considered situation.}
    \label{fig}
\end{figure}

Let's consider an infinite system of positive point charges
along the $x$ axis, separated a distance $a$ from each other. The point charges are evenly distributed on both sides of the coordinate
origin $O$, where a point charge is located, as shown in Figure \ref{fig}.

The electrostatic potential at a distance $z$ from the
origin, on the $+z$ axis, due to a one dimensional lattice ($d=1$ and $y_{n}=0$) is given by the equation (\ref{pott})

\begin{equation}
\label{vv1}
\Tilde{V}_{1}(\xi)=\sum_{n=-\infty}^{\infty} (n^2+\xi^2)^{-1/2},
\end{equation}

\noindent where the sub-index 1 refers to the electrostatic potential in the one dimensional case.

In the above sum we can recognize the Epstein zeta function (\ref{zeta1}) (with $d = 1$), if we  rewrite the sum in terms of a parameter $s$:
\begin{equation}
    \label{z1}
    \zeta_{E}(s,1;\xi^2)=\sum_{n=-\infty}^{\infty} (n^2+\xi^2)^{-s}.
\end{equation}

This infinite sum can be regularized through an analytic continuation of the Epstein zeta function. According to the equation (\ref{zetafinal}), we obtain:

\begin{equation}
    \label{zc1}
    \zeta_{E}(s,1;\xi^2)=\frac{\sqrt{\pi}}{\Gamma(s)}\frac{\Gamma(s-1/2)}{\xi^{2(s-1/2)}}+4\frac{\pi^s}{\xi^{s-1/2}}\sum_{n=1}^{\infty}n^{s-1/2}K_{s-1/2}(2\pi n\xi).
\end{equation}

\noindent Since the term $n=0$ it is excluded from the above sum, we can use the fact that $\sum_{n-\{0\}}f(n)=2\sum_{n=1}^{\infty}f(n)$. However, we must note that if $s=1/2$, the Gamma function in the first term of the equation (\ref{zc1}) has a simple pole. Through change of variable $s=\epsilon+1/2$, we can compute a series expansion as $\epsilon\to 0$, obtaining:

\begin{equation}
    \label{ts}
    \frac{\Gamma(\epsilon)}{\Gamma(1/2+\epsilon)}\frac{1}{\xi^{2\epsilon}}\sim \frac{1}{\sqrt{\pi}}\left(\frac{1}{\epsilon}+2\ln\left(\frac{2}{\xi}\right)\right)+O(\epsilon),
\end{equation}

\noindent neglecting the $O(\epsilon)$ terms, the Epstein zeta function (\ref{zc1}) can be expressed as

\begin{equation}
\label{zc2}
\zeta_{E}(\epsilon,1;\xi^2)=\frac{1}{\epsilon}+2\ln\left(\frac{2}{\xi}\right)+4\sum_{n=1}^{\infty}K_{0}(2\pi n\xi).
\end{equation}

The expression (\ref{zc2}) contains a diverging term, $1/\epsilon$, that  can  be  isolated  by means of the  principal  part prescription \cite{blau}, which means taking the finite part, in order to obtain well defined and finite result for the electrostatic potential in the one-dimensional case

\begin{equation}
    \label{potnorm}
    \Tilde{V}_{1}(\xi)=2\ln\left(\frac{2}{\xi}\right)+4\sum_{n=1}^{\infty}K_{0}(2\pi n\xi).
\end{equation}

Since the Bessel function $K_0(x)$ tends to zero in the
limit of large $x$ (see the Appendix), the asymptotic behavior of the electrostatic potential is given by the dominant first term of equation (\ref{potnorm})

\begin{equation}
    \label{z1ap}
    \Tilde{V}_{1,as}(\xi)\sim 2\ln\left(\frac{2}{\xi}\right).
\end{equation}



As is known, we can obtain the dimensionless electric field form the potential gradient, so, the electrostatic field in the $+z$ axis, in terms of the $\xi$ coordinate, is given by

\begin{equation}
\Tilde{E}(\xi)=-\frac{\partial \Tilde{V}(\xi)}{\partial\xi}.
\end{equation}

Using the general expression for the electrostatic potential (\ref{potnorm}) and noting that

\begin{equation}
\frac{d}{dx}K_{0}(nx)=-nK_{1}(nx),
\end{equation}

\noindent we can obtain an analytic expression for the electric field in the one dimensional case:

\begin{equation}
    \label{e1}
    \Tilde{E}_{1}(\xi)=\frac{2}{\xi}-8\pi\sum_{n=1}^{\infty}n\;K_{1}(2\pi n\xi).
\end{equation}

Therefore, the asymptotic behavior of the electric field in the large $\xi$ regime is given by

\begin{equation}
    \label{e1as}
    \Tilde{E}_{1,as}(\xi)=\frac{2}{\xi}.
\end{equation}



\section{Electrostatic potential and the field of a two-dimensional lattice}

Now, we consider a point charge lattice in the $xy$ plane, as it shows in figure (\ref{lattice}). The electrostatic potential at a distance $z$ of the plane $xy$, on the $+z$ axis, is given by the equation (\ref{pott}). In terms of the Epstein zeta function (\ref{zetafinal}), with $d=2$, we have

 \begin{equation}
     \label{z2d}
     \zeta_{E}(s,2;\xi^2)=\frac{\pi}{\Gamma(s)}\frac{\Gamma(s-1)}{\xi^{2s-2}}+4\frac{\pi^{s}}{\Gamma(s)}\frac{1}{\xi^{s-1}}\sum_{n,m=1}^{\infty}(n^2+m^2)^{(s-1)/2}K_{s-1}\left(2\pi\xi\sqrt{n^2+m^2}\right).
 \end{equation}

\noindent Since the above expression contains a double sum, we have used that $\sum_{n,m-\{0\}}f(n,m)=4\sum_{n,m=1}^{\infty}f(n,m).$

The equation (\ref{z2d}) has no poles when $s=1/2$, so we can evaluate directly to obtain an expression for the electrostatic potential due to this configuration, obtaining the following expression:

\begin{equation}
\label{vt2}
    \Tilde{V}_{2}(\xi)=-2\pi\xi+8\sqrt{\xi}\sum_{n,m=1}^{\infty}\frac{1}{(n^2+m^2)^{1/4}}K_{1/2}\left(2\pi\xi\sqrt{n^2+m^2}\right),
\end{equation}

\noindent where the sub-index 2 refers to the two-dimensional lattice. Note that $K_{-1/2}(x)=K_{1/2}(x)$.

The asymptotic behavior, $\xi\to\infty$, of the electrostatic potential is given by the first term, since the second term in the equation (\ref{vt2}) tends to zero (see the Appendix), so

\begin{equation}
    \label{v2as}
    \Tilde{V}_{2,as}(\xi)\sim -2\pi\xi.
\end{equation}



In order to obtain and analytic expression for the electric field of a two-dimensional lattice, we calculate the potential gradient of equation (\ref{vt2}). By means of equation (\ref{bessel12}), we can express the electric
field for large $\xi$ by

\begin{equation}
    \label{e2}
    \Tilde{E}_{2}(\xi)\approx 2\pi+8\pi\sum_{n,m=1}^{\infty}e^{-2\pi\sqrt{n^2+m^2}\xi},
\end{equation}

\noindent where clearly $e^{-2\pi\sqrt{n^2+m^2}\xi}\to 0$ in the $\xi\to\infty$ regime. So, the asymptotic behavior of the electric field is given by

\begin{equation}
    \label{e2as}
    \Tilde{E}_{2,as}(\xi)=2\pi.
\end{equation}



\section{Conclusions}

We have calculated the
electrostatic potential and the field of one and two-dimensional infinite
lattice of point charges, using the zeta function regularization method. Through the analytic extension of the Epstein zeta function, we obtained expressions that only in the one-dimensional case was necessary to remove $1/\epsilon$ singularity using the principal prescription part. In the two-dimensional case, no such singularity appeared, and the resulting expressions become very useful for numerical evaluations. We explored the asymptotic behavior of the electrostatic potential and field in one and two-dimensional distributions. In the one-dimensional case, we found that the dominant terms of the potential and field behave like an infinite charged line, where $V(r)\sim\ln(1 / r)$ and $E(r)\sim 1/r$, respectively. On the other hand, for the two-dimensional case, the asymptotic behavior of the electrostatic potential and field, are similar to the infinite charge plate, where $V(r)\sim -r$ and $E(r)= \text{constant}$.

Finally, this example shows some features of the zeta function regularization
applied to an electrostatic problem, which can be useful as a pedagogical
exercise for graduate students studying spectral functions

\section*{Acknowledgment}

The author thanks Professors Julio M Yáñez, Juan C Rojas and Roberto A Lineros (Universidad Católica del Norte, Antofagasta, Chile) for valuable suggestions and time spent in many discussions.

\section*{Appendix}

The asymptotic expansion of the Bessel function of second kind is given by \cite{abramowitz}

\begin{equation}
    \label{besselexpansion1}
    K_{\nu}(x)\sim\sqrt{\frac{\pi}{2x}}e^{-x}
\end{equation}

\noindent For $x\to\infty$ with $\nu$ fixed. In the special case $\nu=1/2$ we have

\begin{equation}
    \label{bessel12}
    K_{1/2}(x)=\sqrt{\frac{\pi}{2x}}e^{-x}.
\end{equation}

The Bessel function of second kind obtained in equations  (\ref{potnorm}), (\ref{vt2}), (\ref{e1}) and (\ref{e2}),  has the form

\begin{equation}
    \label{besselexpansion2}
  f(x)=\sum_{|\textbf{n}|\in\mathbb{Z}^{d}-\{0\}}K_{\nu}(|\textbf{n}| x)\sim\sum_{|\textbf{n}|\in\mathbb{Z}^{d}-\{0\}}\sqrt{\frac{\pi}{2|\textbf{n}|x}}e^{-|\textbf{n}|x}
\end{equation}

\noindent where $|\textbf{n}|=\sqrt{n_1^2+n_2^2+\cdots+n_d^2}$ represents a sequence of real numbers. Since we are interested in the asymptotic behavior of $f(x)$, this expression is very useful for
numerical evaluations. However, in the $x\to\infty$ regime and considering that $|\textbf{n}|$ is an increasing sequence of numbers, we can see that the numerical value of $f(x)$ goes to zero. So we can consider that $f(x)$ in $x\to\infty$ regime can not be considered as a dominant term.

\subsection{List of symbols}

\centering
\begin{tabular}{ll}
$\rea$ &  set of all real numbers\\
$\mathbb{Z}$ & set of all integers (whether positive, negative or zero)\\
$\mathbb{C}$ & set of all complex numbers\\
$\mathbb{N}$ & set of all natural numbers\\
$\mathbb{N}_{0}$ & set of all natural numbers including zero\\
$\in$ & is member of\\
$\{\;\}$ & set\\
$\epsilon_0$ & permittivity of free space\\
$K_{\nu}(x)$ & Modified Bessel function of second kind\\
$\Gamma(x)$ & Gamma function
\end{tabular}



\end{document}